\documentclass[floatfix, twocolumn]{article}

\usepackage{graphicx}
\usepackage{amsmath,amssymb,amsfonts}
\usepackage{color}
\usepackage{hyperref}
\usepackage{textcomp}
\usepackage{verbatim}
\usepackage[numbers]{natbib}

\date{October, 2013} 

\title{System analysis of Force Feedback Microscopy}

\author{Mario S. Rodrigues$^{1*}$, Luca Costa$^{2,3}$, Jo\"{e}l Chevrier$^{2,3,4}$ and Fabio Comin$^2$ \\
\small{1- CFMC/Dep. de F\'{\i}sica, Universidade de Lisboa,Campo Grande 1749-016 Lisboa, Portugal} \\
\small{2- European Synchrotron Radiation Facility, 6 rue Jules Horowitz BP 220, 38043 Grenoble Cedex, France} \\
\small{3- Universit\'{e} Grenoble Alpes, Inst NEEL, F-38042 Grenoble, France} \\
\small{4- CNRS, Inst NEEL, F-38042 Grenoble, France} \\
}

\begin{document}

\maketitle

\begin{abstract}
It was shown recently that the Force Feedback Microscope
can avoid the jump-to-contact in Atomic force Microscopy even when the 
cantilevers used are very soft, thus increasing force resolution.
In this letter, we explore theoretical aspects of the associated real time 
control of the tip position. We take into account lever parameters such as the 
lever characteristics in its environment, spring constant, mass, dissipation 
coefficient and the operating conditions such as controller gains, and 
interaction force. 
We show how the controller parameters are determined so that the FFM 
functions at its best and estimate the bandwidth of the system under these 
conditions.
\end{abstract}

\let\thefootnote\relax\footnote{$^*$ Corresponding author: \textit{mmrodrigues@fc.ul.pt}}

\section{Introduction}

The Atomic Force Microscope (AFM) was introduced almost 30 years ago
\cite{Binnig1986}.
The idea consists in mounting an ultra sharp tip on a beam (cantilever)
and then scan it over a surface while 
recording the deflection of the beam.
Since its invention, the technique has been and still is progressing
very fast and with it our understanding of phenomena that occur at the 
nanoscale and sometimes even at the atomic level.
It naturally evolved from a static to a dynamic technique \cite{martin87a}.
Dynamic AFM, in turn, allowed for other techniques to emerge, 
such as magnetic force microscopy \cite{martin87b}, 
electrostatic force microscopy \cite{martin88},
kelvin probe microscopy \cite{nonnenmacher91}
and many other modes have been introduced since then.
Yet, despite the enormous evolution the AFM has made,
it still has the limitation that the tip position 
becomes unstable at close proximity to the sample.
This instability, commonly referred to as \textit{jump-to-contact},
happens when the cantilevers used have spring constants on the order of the 
N/m comparable to the force gradients that form when the tip is brought close to 
the sample surface.
When the attractive tip-sample force gradient equals the spring constant of the 
cantilever the instability occurs.
Avoiding the jump-to-contact implies a minimum limit either in the 
cantilever stiffness or in the oscillation amplitude. 
For a particular interaction, if increasing the cantilever stiffness is not 
sufficient to overcome the tip instability, then increasing enough its kinetic 
energy avoids the jump to contact \cite{giessibl97}.
These strategies however pose several problems: 
the stiffness of the cantilever limits the resolution in force; large 
amplitudes of oscillation make it difficult to quantitatively analyze 
the interaction and the later also decrease the lateral 
resolution \cite{giessibl03}.
The use of very stiff cantilevers, such as tuning forks \cite{giessibl03} solve 
the instability problem but make it impossible to measure the tip-sample forces.
To overcome this conundrum some solutions have been proposed in the past 
\cite{jarvis_96, Ashby2000, Goertz2010} and more recently Force Feedback 
Microscopy (FFM) \cite{io12, jmr13eu}.
These techniques have in common the use of a feedback loop to 
maintain tip stability by counteracting the 
tip sample force with an equal but opposite amount of force.

Here we will consider the case where the counteracting force is controlled 
by a proportional, integral and differential controller (PID).
The force results from the addition of three components:
proportional to the tip position $g_p$; 
proportional to the time integral of the tip position $g_i$; 
and proportional to the time derivative of the tip position $g_d$.
the PID controller gains are $g_p$, $g_i$ and $g_d$ respectively.
Hereafter, we shall refer to this force as $F_{pid}$.
In this letter, we  work on some of the theoretical aspects of the technique.
First we present the model system and study the conditions within which 
this system is stable.
We then give an estimate of the expected bandwidth.
However, we do not attempt to estimate the maximum permitted 
proportional gain \cite{Kato1999}.
Finally we study how the implementation of this strategy affects the 
harmonic modes of the oscillator and how the interactions can be calculated from 
amplitude and phase changes using this type of control.

\section{The model system}

The cantilever plus tip are considered as a mass-spring system,
with spring constant $k$, mass $m$ and with some damping $\gamma$.
In our approach we assume the mass is subjected to some tip sample force 
$F_{ts}$ and to the control force $F_{pid}$.
The motion of the mass is described by the equation below:
\begin{equation}
 m \ddot{x} = F_{ts} + F_{pid}-k x - \gamma \dot{x} 
\end{equation}
Where the control force depends on the tip 
position in the following way:
\begin{equation}
F_{pid}= -g_p x - g_d \dot{x} - g_i \int x dt 
\label{eq:2}
\end{equation}
This force is directly determined by the real time action of the controller on 
the piezo element at the cantilever base.
From here on, we concentrate only on the cases were
the tip-sample interaction force can be expressed as:
\begin{equation}
F_{ts}=F_{ts,0}-k_{ts} x-\gamma_{ts} \dot{x}
 \label{eq:ukn}
\end{equation}
$F_{ts,0}$ is a term that does not depend on the position, $k_{ts} x$ is 
an elastic term proportional to the tip position
and $\gamma_{ts} \dot{x}$ is a damping term proportional to the tip speed.
In this case the total stiffness and damping coefficient can be written as 
$k_t=k+k_{ts}$ and $\gamma_t=\gamma+\gamma_{ts}$ respectively.
The equation of motion of such system 
can be described using Laplace transforms as follows:
\begin{equation}
 m X s^2 +(k_t+g_p) X +(\gamma_t + g_d) X s + \frac{g_i}{s} X=\mathcal{L}\{F_{ts,0}\}
 \label{eq:0}
\end{equation}
hence the tip position:
\begin{equation}
 X=\frac{s \mathcal{L}\{F_{ts,0}\}}{m s^3+(\gamma_t + g_d) s^2+ (k_t + g_p)s + g_i}
 \label{eq:lap}
\end{equation}
This solution is that of an harmonic oscillator when the integral gain $g_i=0$.
Next, we propose to evaluate this expression when 
the system is submitted to an impulse to check its stability
and then how it responds to harmonic stimuli.

\section{Stability criteria}

Let us take the case were $F_{ts,0}=0$ for $t<0$ and $F_{ts,0}=F_0$ for 
$t \geq 0$, in which case the Laplace transform is simply $F_0/s$.
Analyzing this response allows us to obtain the conditions within which 
the system is stable.
The behavior of the system will depend on the roots of the denominator in 
equation \ref{eq:lap}. It is a third order polynomial so the zeros are better 
found numerically.
Nevertheless, we can approximate the denominator by another polynomial such 
that the zeros are easy to calculate.
We can separate this in two different regimes, the under-damped regime and the over-damped regime.
The first regime would correspond to the situation where the AFM is operated in air or vacuum, whereas the second regime 
is more likely to occur when the cantilever motion is damped by liquid.

\subsection{Under-damped case}

In the under-damped case the denominator of equation \ref{eq:lap} has one 
real root and the two other roots are complex conjugates.
The complex conjugates cause the system to oscillate
whereas the real root is responsible for changing the equilibrium position
around which the system oscillates.
\begin{equation}
x(t)=\frac{F_0}{k} \frac{\omega_0^2 e^{-\lambda_0 t}+ \alpha \omega_0^2 
e^{-\lambda_c t} \cos(\omega_c t+\phi)}{\omega_c^2+(\lambda_0-\lambda_c)^2 }
\label{eq:dyn1}
\end{equation}
This is simply the inverse Laplace transform of equation \ref{eq:lap}
that is the solution to equation 1. 
The system decays to some mean value within a time $1/\lambda_0$ and oscillates
with frequency $\omega_c$ with an amplitude that decays with a time constant $1/\lambda_c$.
For simplicity we will not explicit the values of 
constants $\alpha$ or $\phi$ that do not matter for the purpose of discussing 
the stability of the system.
Here $\lambda_0$ is the real root, $\lambda_c$ is the real part of the 
complex root, $\omega_c$ its respective imaginary part and 
$\omega_0=\sqrt{k/m}$ is the natural frequency of the system.
Note that the denominator of equation \ref{eq:lap} can be rewritten as:
\begin{equation}
 \left(s+\frac{g_i}{k_t+g_p} \right)(m s^2+a s+b)+c
\end{equation}
\noindent where
\begin{eqnarray}
 a &= \gamma_t + g_d-\frac{m g_i}{k_t+g_p}, &b=k_t+g_p-a\frac{g_i}{k_t+g_p},
\nonumber \\
 c &=g_i \left(1-\frac{b}{k_t+g_p} \right). &
\end{eqnarray}
In the under-damped case $a$ is very small.
If we replace $a$ by zero in the expression of $b$ we see that indeed
$c=0$. Neglecting $c$, allows to easily find the three roots.
\begin{eqnarray}
 \lambda_0 &=& \frac{g_i}{k_t+g_p} \nonumber \\ 
 \lambda_c &=& \frac{\omega_0^2 (\gamma_t+g_d)}{2 k}-\frac{g_i}{2(k_t+g_p)} 
\nonumber \\
\omega_c &=& \omega_0 \sqrt{\frac{k_t+g_p}{k}-\left(\frac{(\gamma_t+g_d) 
\omega_0}{2k} 
-\frac{g_i/\omega_0}{2 (k_t+g_p)}\right)^2}
 \end{eqnarray}
If either $\lambda_0$ or $\lambda_c$ are negative then $x$ diverges.
The conditions in which they are positive give us the stability criteria:
\begin{eqnarray}
 \textrm{crit1a}: \hspace{1cm} & \lambda_0>0 \Rightarrow k_t+g_p > 0 \nonumber 
\\
 \textrm{crit2a}: \hspace{1cm} & \lambda_c>0 \Rightarrow g_i< \omega_0^2 
(\gamma_t+g_d) \frac{k_t+g_p}{k}
\end{eqnarray}
The first criterion relates to the \textit{jump to contact}. 
In conventional AFM it happens when $ k_t = 0$ that occurs when the tip-sample 
force gradient equals the cantilever spring constant.
In FFM this can be avoided trough the use of the proportional gain $g_p$.
The second criterion imposes a superior limit 
to the integral gain $g_i$ and thus limits the bandwidth of operation of 
the FFM.
If criterion 2 is not met, then the system will oscillate with
ever increasing amplitude.
Criterion 2 is not a useful criterion because 
for integral gains approaching that limit,
the time it takes the oscillator to reach stability 
approaches infinity.
As a matter of fact, ideally we want to maximize both $\lambda_0$ and 
$\lambda_c$, corresponding to a restore of the equilibrium position and 
amplitude of oscillation to zero as fast as possible without instability.
The integral gain $g_i$ that corresponds to this situation is:
\begin{eqnarray}
 \textrm{crit3a}: \hspace{1cm} & g_i = \omega_0^2 (\gamma_t+g_d) 
\frac{k_t+g_p}{3 k}
\end{eqnarray}
\begin{figure}[ht]
 \includegraphics[width=3.33in]{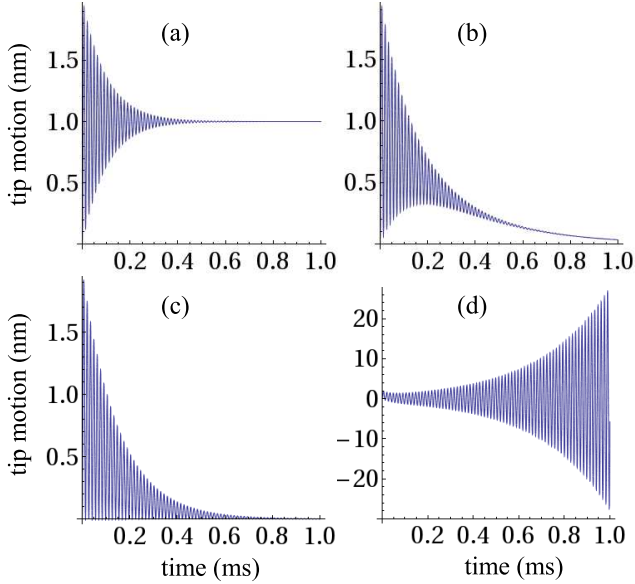}
 \caption{Tip motion after a step function: (a) without integral gain $g_i$
 (b) integral gain equal to 1/2 of the ideal integral gain, (c) ideal integral 
gain (6667 N/ms) and  (d) integral gain 4 times the ideal integral gain. 
Cantilever parameters are $k=1$ N/m, $f_0 = $70kHz and $\gamma=10^{-7}$ kg/s.}
\end{figure}
The second and third criteria show how the 
use of a proportional gain increases the maximum
integral gain before the system becomes unstable and 
how the its maximum value depends on the
tip-sample interaction i.e. if $k_t+g_p \approx 0$ 
then the maximum integral approximates zero
thus reducing the bandwidth of the system.
Fig. 1 shows the behavior of the system for four different integral gains.
When the integral gain is zero one recognizes the behavior of a weekly damped 
oscillator. When $g_i>0$ the equilibrium position of the system can be seen as 
changing in time with a decay length $1/\lambda_0$. The frequency of the 
oscillation $\omega_c$ is in all cases very approximated to $\omega_0$ which is 
natural of a weakly damped oscillator.
The data shown in Fig. 1 is relative to a numerical calculation of the 
roots. The approximations used here were also calculated and yielded values 
that that are precise to more than 1/1000. 
\subsection{Over-damped case}
In this case the roots are all real for moderate PID gains.
Note however that despite the system being 
over-damped, a large enough integral 
gain can bring the system into oscillation 
in which case two of the roots will be complex conjugates.
This is however one situation that is to be avoided as there is no 
advantage in that situation. Unlike in the 
under-damped case, if the system exhibits oscillations,
these do not occur close to the natural frequency of the oscillator as 
they are driven by the controller gains rather than by 
the dynamics of the cantilever.
For moderate gains, we can neglect the term
on $s^3$ and it becomes straightforward to calculate the roots.
The response of the system is in this case:
\begin{equation}
 x(t)=\frac{F_0 (e^{-(\lambda_1-\lambda_2)t}-e^{-(\lambda_1+\lambda_2)t})}{2 \gamma_t \lambda_2}
\end{equation}
where:
\begin{eqnarray}
 \lambda_1 &=& \frac{k_t+g_p}{2(\gamma_t+g_d)} \nonumber \\
 \lambda_2 &=& \sqrt{\lambda_1^2- g_i/(\gamma_t+g_d)}
\end{eqnarray}
Thus, to avoid oscillations it must be:
\begin{equation}
 (k_t + g_p)^2 > 4 (\gamma_t+g_d) g_i
\label{eq:cond}
\end{equation}
In which case $\lambda_1$ must always be positive otherwise the system diverges.
If $\lambda_1$ is negative then $\lambda_2$ would have to be negative 
for the system not to diverge. The condition that $\lambda_2$ is negative is 
physically not possible. Therefore one first criterion for stability must be:
\begin{eqnarray}
 crit1b: \hspace{1cm} & k_t+g_p > 0
\end{eqnarray}
Note that this criterion is the same as criterion 1a we found before.
Depending on condition \ref{eq:cond} the system may or not oscillate and these 
oscillations may or not decay to zero.
It can be shown (figure 2) that an integral gain $g_i$ slightly above the one corresponding to 
equality in condition \ref{eq:cond}
produces negligible oscillation while driving the equilibrium position to zero 
faster. A useful criterion is:
\begin{eqnarray}
 crit2b: \hspace{1cm} & g_i \lesssim \frac{(k_t+g_p)^2}{\gamma_t+g_d}
\end{eqnarray}
Increasing the integral above this value only increases 
the oscillations without gain in performance (see Fig. 2), eventually 
leading to instabilities.
Criterion 2b represents the gain that restores the equilibrium position to 
zero, in about the minimum amount of time with negligible ringing.
To visualize the behavior more intuitively let us take the case where $g_i$ is well below that limit:
  \begin{eqnarray}
    \lambda_1-\lambda_2 \approx \lambda_0 = \frac{g_i}{k_t+g_p} \nonumber \\
  \lambda_2 + \lambda_1 \approx 2 \lambda_1
\end{eqnarray}
Then equation of motion becomes:
\begin{equation}
 x(t) \approx \frac{F_0 (e^{-\lambda_0 t}-e^{-2 \lambda_1t})}{k_t+g_p}
\end{equation}
Thus when the tip is subjected to a step force the position
will change exponentially away from its initial position and then come back
to the initial position.

\begin{figure}[h]
 \includegraphics[width=3.33in]{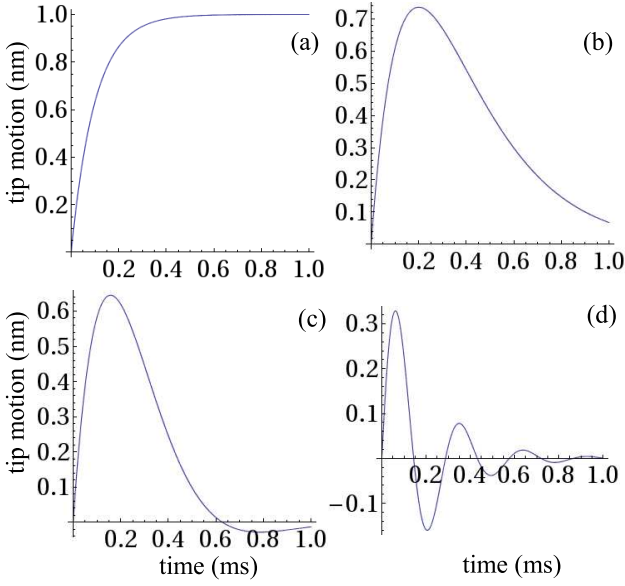}
 \caption{Tip motion after a step function: (a) without integral gain $g_i$,
 (b) integral gain corresponding to equality in condition \ref{eq:cond} 
(2500N/ms),(c) corresponding to crit. 2b and (d) 10 times the previous gain. 
Cantilever parameters are $k=1$N/m, $f_0 = $70kHz and $\gamma=10^{-4}$kg/s.}
\end{figure}
Fig. 2 shows the behavior of the system for four different integral gains.
When the integral gain is zero one recognizes the behavior of a strongly damped 
oscillator. When $g_i>0$ below the critical value, the equilibrium 
position of the system can be seen as decaying with a decay length 
of $1/\lambda_0$. For larger gains the system starts 
exhibiting oscillations.

\section{General criteria}

Here we introduce the quality factor $Q$ of the system, a commonly used 
parameter in AFM:
\begin{equation}
 Q \equiv \frac{k}{\omega_0 \gamma}
 \label{eq:Q}
\end{equation}
To simplify the discussion, let us consider the case where both the differential gain 
and the tip-sample dissipation are small.
Note that this will lead to lower limits for maximum integral gains.
In that case the criteria for stability are:
\begin{eqnarray}
g_p & > -(k+k_{ts})  \\
g_{i,under,max} & = \frac{\omega_0}{3 Q} (k_t+g_p) & \textrm{under-damped}  \\
g_{i,over, max} & =  Q \omega_0 \frac{(k_t+g_p)^2}{k}  & \textrm{over-damped}
\end{eqnarray}
We can include the effect of the differential gain by replacing the $Q$ factor 
by an effective Q factor in which the differential gain is added to $\gamma$ in 
equation eq. \ref{eq:Q}. The same procedure can be done to include the sample 
dissipation.
One of the consequences of the result above is that if the tip-sample force 
gradient ($k_{ts}$) equals or is smaller than the negative of the cantilever 
stiffness, the jump to contact can still be avoided provided the
proportional gain is large enough.
For an under-damped oscillator smaller $Q$ factors are 
more favorable whereas the contrary is true for an over-damped oscillator.
For a given cantilever frequency and stiffness the best 
situation for the FFM operation is when the cantilever motion is close to 
the critical-damped regime.
In the under-damped case the equilibrium position decays to zero 
with a time constant of $\tau=1/\lambda_0$, hence $\tau=(k_t+g_p)/gi$.
Notice the same is also approximately true in the over-damped case.
The bandwidth of the FFM can be estimated from these limits:
\begin{eqnarray}
\omega_a = &\frac{\omega_0}{3 Q} \hspace{1cm} & \textrm{under-damped}\\
\omega_b = &Q \omega_0 \frac{k_t+g_p}{k} \hspace{1cm} & \textrm{over-damped}
\end{eqnarray}
The ideal cantilever is one with high resonance frequency 
and close to critical damped.
An example of one such type of cantilever would be 
the ones used in high speed AFM \cite{ando08}.

\section{Maximum approach speed}
Let us now consider an approach curve experiment.
If a unit  step force $F_0$ is applied to the system the maximum change of 
tip position $\Delta x$ is:
\begin{equation}
 \Delta x=\frac{2 \Delta F}{(k_t+g_p)}
\end{equation}
The factor 2 is for a weakly damped system whereas for a damped system this 
factor is less than one. 
To have a total tip motion $x$ never greater than this value above
the system must be allowed to relax a time
$\/\tau$ before another step of magnitude $\Delta F (1-1/e) \approx 0.63 
\Delta F$ can be applied. Here $e$ is the base of the natural logarithm. This 
is:
\begin{equation}
 \Delta F<0.63 \frac{k_t+g_p}{2} \Delta x
\end{equation}
In an approach curve $\Delta F$ can be put as $\Delta z k_{ts}$,
(how much the sample is approached times the spring constant of the 
interaction) and $\Delta z$ can be divided by $\tau$ to give a velocity:
\begin{equation}
 v<0.63 \frac{k_t+g_p}{2 k_{ts} \tau} \Delta x
\end{equation}
This is the maximum speed at which an approach curve can be taken.
Taking the time constant $\tau=1/\lambda_0$ and remembering $\lambda_0$
gives:
\begin{equation}
 v<0.63 \frac{g_i \Delta x}{2 k_{ts}} 
\end{equation}
The maximum speed depends on the integral gain and on the tip sample 
interaction.
The maximum integral gain in turn depends on all the other constants defining the 
system. If we replace $g_i$ by its maximum values that assure stability we find:
\begin{eqnarray}
 v <0.63 \omega_a \frac{\Delta x}{2} \frac{(k_t+g_p)}{k_{ts}} & 
\hspace{1cm} \textrm{under} \\
 v<0.63 \omega_b \frac{\Delta x}{2}\frac{(k_t+g_p)^2}{k k_{ts}}  & 
\hspace{1cm} \textrm{over}
\end{eqnarray}

The proportional gain plays a crucial role.
Not only it must be such that $(k+k_{ts}+g_p)>0$ to remove the instability,
but it also limits the speed that becomes zero when the sum above is not some 
limited value above zero.
Experimental set ups that cannot provide an instantaneous force proportional to 
the position of the tip will not be able to overcome the jump to contact.

As an example, take the limiting case were $k_{ts}=-1$N/m, $k=1$N/m and $g_p=1$N/m.
If we use the same limits above and accept a displacement of the tip of $\Delta 
x=-0.1nm$ we obtain a maximum speed of 105 nm/s.

We have simulated an approach curve experiment in the limit where 
jump-to-contact usually occurs. Four cases were considered: no PID control; only 
integral control; integral and proportional control.
For the later we have considered an approach curve at the maximum speed 
estimated above, and at half that speed (52nm/s).
Fig 3. shows the PID force and the tip position for these cases. 
\begin{figure}[h]
 \includegraphics[width=3.33in]{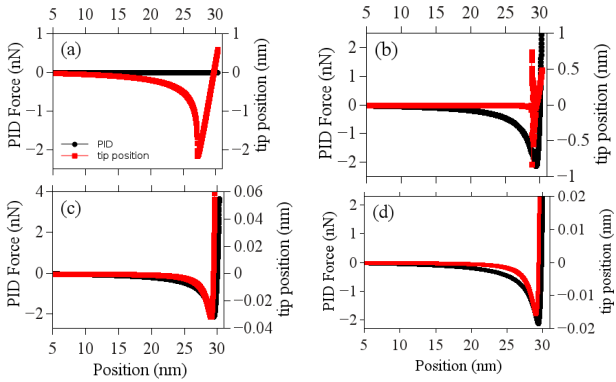}
 \caption{Without PID control (a) one notices the jump to contact. 
With the integral gain alone (b) there is no jump-to-contact and the force 
is correctly measured by the PID although the tip position is unstable and may 
occasionally tap the surface. With integral and proportional control (c, d) the 
tip instabilities can be completely removed.
The cases (c) and (d) are relative to an approach speed of 105 and 53 nm/s respectively.
}
\end{figure}
Fig. 3 shows that the error is smaller than the one estimated here. 
This is because most of the time the interaction is smaller than the value 
used for estimating the maximum speed.
For the simulation we have used a force of the type $F_{ts}=A/z+B/z^2$ where 
the coefficients $A$ and $B$ were chosen so that the maximum value of $\partial 
F/\partial z$ was 1 N/m hence equal to the cantilever spring constant.
\section{Harmonic response}

We can analyze the frequency behavior of the system by exciting it with a 
Dirac delta, in which case the Laplace transform of the motion of the system is:
\begin{equation}
X = \frac{F_\omega}
	  {m s^2+g_p+k_t+(\gamma_t+g_d)s+g_i/s}
 \label{eq:Ftotal}
 \end{equation}
The amplitude of the response to an harmonic 
excitation and the respective phase are given by:
%
\begin{equation}
 R = \frac{F_\omega}{\sqrt{[(k_t+g_p)-m \omega^2]^2+[(\gamma_t + g_d) \omega-\frac{g_i}{\omega}]^2}}
\label{eq:R}
 \end{equation},
\begin{equation}
 \phi = \arctan\left[\frac{(\gamma_t + g_d)\omega-\frac{g_i}{\omega}}{(k_t + g_p)-m \omega^2}\right]
\label{eq:phi3}
 \end{equation}
 %
 %
The response of the oscillator obviously depends on the PID gains.
However we are interested in the force/response ratio $F_r$.
For that we must compute the sum of the forces $F_{sum}$ exciting the 
cantilever, which is the excitation plus the PID response to that.
\begin{equation}
F_{sum} = F_\omega-\frac{F_\omega (g_p+g_i/s+g_ds)}
	  {m s^2+g_p+k_t+(\gamma_t+g_d)s+g_i/s}
 \label{eq:Fsum}
 \end{equation}
The ratio $F_{sum}/X$ is:
\begin{equation}
\frac{F_{sum}}{X} = k_t +\gamma_t s + m s^2 
\label{eq:FsumX}
\end{equation},
Naturally this ratio does not depend on the PID gains.
But note that to measure this ratio it implies to measure the total 
excitation force and not just the harmonic supplied stimulus.
The absolute value of the previous quantity is:
\begin{equation}
F_r = \sqrt{(k_t-m \omega^2)^2+\gamma_t^2 \omega^2} 
\label{eq:Fr}
\end{equation},
The phase difference between $F_{sum}$ and $X$ also does not depend on the 
PID gains and is given by
\begin{equation}
\phi = \arctan\left(\frac{\gamma_t \omega}{k_t - m \omega^2} \right)
\label{eq:phi}
\end{equation},
From the two previous equations it follows:
\begin{equation}
F_r \cos(\phi) = k_t-m \omega^2, \hspace{1cm} F_r \sin(\phi) = \gamma_t \omega
\label{eq:phi2}
\end{equation}
To proceed, we insert the information known from when there are no tip-sample 
forces, and we identify the respective amplitude and phase with the the 
superscript $0$. In that case is:
\begin{equation}
F_r^0 \cos(\phi^0) = k-m \omega^2, \hspace{1cm} 
F_r^0 \sin(\phi^0) = \gamma \omega
\end{equation}
\noindent Combining the last four equations yields the following final 
expressions:
\begin{equation}
k_{ts}= F_r^0 \left[n \cos(\phi)-\cos(\phi_0)\right]
\label{eq:km}
\end{equation}
\begin{equation}
 \gamma_{ts}=\frac{F_r^0}{\omega} \left[n \sin(\phi)-\sin(\phi_0)\right]
\label{eq:gm}
 \end{equation}
\noindent where $F_r^0$ is the force/response ratio
of the unperturbed oscillator and $n=F_r/F_r^0$ is the normalized 
force/response ratio.
In a measurement, $n$ and $\phi$ can be easily measured.
When using FFM one question that may arise is that of the sensitivity:
is the sensitivity given by the cantilever properties or by the effective
cantilever properties that are changed by the PID?
The answer is that it depends on how the excitation is taken into 
account.
Notice that the PID contributes to excite the cantilever (\ref{eq:Fsum}) and 
that contribution contains information about the tip-sample interaction.
If the total excitation is measured, then the sensitivity is intrinsically
given by the cantilever properties, whereas if only the harmonic stimulus
$F_\omega$ is measured, then the sensitivity is given by the effective 
cantilever parameters. 
If the PID gains are moderate, then the dynamic response of the 
cantilever is not too much affected, in which case to compute the interaction 
as a function of $F_{sum}$ or $F_\omega$ yields the same result.

As a conclusion to this section, the sensitivity in dynamic mode depends on the 
spring constant of the cantilever in the same way as in conventional AFM, but 
the spring constants required to avoid the jump to contact in FFM are 
lower than those required for conventional AFM.

\section{Conclusions}

In conclusion we show that with the use of a PID feedback loop
to control the tip position it is no longer required to use cantilevers with 
spring constants larger than the tip-surface force gradient
to avoid the jump-to-contact as long as a large enough proportional actuation 
is done to the tip effectively changing the cantilever spring constant to 
$(k+g_p)$.
The fact that $(k_t+g_p)$ never is zero or negative means that the cantilever 
always has a limited valued equilibrium position, and is a requirement for FFM 
to work.
The proportional gain is also relevant to increase the maximum integral gain
that can be applied to the system without causing instabilities, thus 
increasing the bandwidth of the technique.
The integral gain $g_i$ will work to maintain the equilibrium position of the 
tip at the same place.
Finally, we conclude that the use of the proportional gain effectively 
increases the cantilever spring constant but this effective augmentation of the 
cantilever spring constant does not result in loss of sensitivity.

\section*{Aknowledgments}
Mario S. Rodrigues acknowledges financial support from Funda\c{c}\~{a}o
para a Ci\^{e}ncia e Tecnologia SFRH/BPD/69201/2010. Luca Costa acknowledges COST Action TD 1002.

\bibliographystyle{abbrvnat}


\begin{thebibliography}{14}
\providecommand{\natexlab}[1]{#1}
\providecommand{\url}[1]{\texttt{#1}}
\expandafter\ifx\csname urlstyle\endcsname\relax
  \providecommand{\doi}[1]{doi: #1}\else
  \providecommand{\doi}{doi: \begingroup \urlstyle{rm}\Url}\fi

\bibitem[Ando et~al.(2008)Ando, Uchihashi, Kodera, Yamamoto, Miyagi, Taniguchi,
  and Yamashita]{ando08}
T.~Ando, T.~Uchihashi, N.~Kodera, D.~Yamamoto, A.~Miyagi, M.~Taniguchi, and
  H.~Yamashita.
\newblock High-speed afm and nano-visualization of biomolecular processes.
\newblock \emph{Eur J Physiol}, 456:\penalty0 211--225, 2008.

\bibitem[Ashby et~al.(2000)Ashby, Chen, and Lieber]{Ashby2000}
P.~D. Ashby, L.~Chen, and C.~M. Lieber.
\newblock Probing intermolecular forces and potentials with magnetic feedback
  chemical force microscopy.
\newblock \emph{Journal of the American Chemical Society}, 122\penalty0
  (39):\penalty0 9467--9472, 2000.

\bibitem[Binnig et~al.(1986)Binnig, Quate, and Gerber]{Binnig1986}
G.~K. Binnig, C.~F. Quate, and C.~Gerber.
\newblock {Atomic Force Microscope}.
\newblock \emph{Phys. Rev. Lett.}, 56:\penalty0 930, 1986.

\bibitem[Costa et~al.(2013)Costa, Rodrigues, Newman, Zubieta, Chevrier, and
  Comin]{jmr13eu}
L.~Costa, M.~S. Rodrigues, E.~Newman, C.~Zubieta, J.~Chevrier, and F.~Comin.
\newblock Imaging material properties of biological samples with a force
  feedback microscope.
\newblock \emph{J. Mol. Recognit.}, ??\penalty0 (1–4):\penalty0 ???, 2013.

\bibitem[Giessibl(1997)]{giessibl97}
F.~J. Giessibl.
\newblock Forces and frequency shifts in atomic-resolution dynamic-force
  microscopy.
\newblock \emph{Phys. Rev. B}, 56:\penalty0 16010--16015, Dec 1997.

\bibitem[Giessibl(2003)]{giessibl03}
F.~J. Giessibl.
\newblock Advances in atomic force microscopy.
\newblock \emph{Rev. Mod. Phys.}, 75:\penalty0 949--983, Jul 2003.

\bibitem[Goertz and Moore(2010)]{Goertz2010}
M.~Goertz and N.~Moore.
\newblock Mechanics of soft interfaces studied with displacement-controlled
  scanning force microscopy.
\newblock \emph{Progress in Surface Science}, 85\penalty0 (9–12):\penalty0
  347 -- 397, 2010.

\bibitem[Jarvis et~al.(1996)Jarvis, Yamada, Yamamoto, Tokumoto, and
  Pethica]{jarvis_96}
S.~Jarvis, H.~Yamada, S.-I. Yamamoto, H.~Tokumoto, and J.~Pethica.
\newblock Direct mechanical measurement of interatomic potentials.
\newblock \emph{Nature}, 384:\penalty0 247--249, Nov. 1996.

\bibitem[Kato et~al.(1999)Kato, Kikuta, Nakano, Matsumoto, and Iwata]{Kato1999}
N.~Kato, H.~Kikuta, T.~Nakano, T.~Matsumoto, and K.~Iwata.
\newblock System analysis of the force-feedback method for force curve
  measurements.
\newblock \emph{Review of Scientific Instruments}, 70\penalty0 (5):\penalty0
  2402--2407, 1999.

\bibitem[Martin and Wickramasinghe(1987)]{martin87b}
Y.~Martin and H.~K. Wickramasinghe.
\newblock Magnetic imaging by "force microscopy" with 1000 a resolution.
\newblock \emph{Appl. Phys. Lett.}, 50:\penalty0 1455--1457, 1987.

\bibitem[Martin et~al.(1987)Martin, Williams, and Wickramasinghe]{martin87a}
Y.~Martin, C.~C. Williams, and H.~K. Wickramasinghe.
\newblock Atomic force microscope-force mapping and profiling on a sub 100-a
  scale.
\newblock \emph{J. Appl. Phys.}, 61:\penalty0 4723--4729, 1987.

\bibitem[Martin et~al.(1988)Martin, Abraham, and Wickramasinghe]{martin88}
Y.~Martin, D.~W. Abraham, and H.~K. Wickramasinghe.
\newblock High-resolution capacitance measurement and potentiometry by force
  microscopy.
\newblock \emph{Appl. Phys. Lett.}, 52:\penalty0 1103--1105, 1988.

\bibitem[Nonnenmacher et~al.(1991)Nonnenmacher, O'Boyle, and
  Wickramasinghe]{nonnenmacher91}
M.~Nonnenmacher, M.~P. O'Boyle, and H.~K. Wickramasinghe.
\newblock Kelvin probe force microscopy.
\newblock \emph{Applied Physics Letters}, 58\penalty0 (25):\penalty0
  2921--2923, 1991.

\bibitem[Rodrigues et~al.(2012)Rodrigues, Costa, Chevrier, and Comin]{io12}
M.~S. Rodrigues, L.~Costa, J.~Chevrier, and F.~Comin.
\newblock Why do atomic force microscopy force curves still exhibit jump to
  contact?
\newblock \emph{Applied Physics Letters}, 101\penalty0 (20):\penalty0 203105,
  2012.

\end{thebibliography}

\end{document}